\documentclass[prl,aps,floatfix,showpacs,tightenlines,twocolumn,superscriptaddress,amsmath,amssymb]{revtex4}

\usepackage{graphicx}
\usepackage{dcolumn}
\usepackage{bm}
\usepackage{epsfig}

\begin{document}

\title{Hybrid Quantum Repeater Using Bright Coherent Light}

\author{P.\ van Loock}\email{vanloock@nii.ac.jp}
\affiliation{National Institute of Informatics, 2-1-2
Hitotsubashi, Chiyoda-ku, Tokyo 101-8430, Japan}

\author{T.\ D.\ Ladd}
\affiliation{Edward L. Ginzton Laboratory, Stanford University,
Stanford, California 94305-4088, USA} \affiliation{Nanoelectronics
Collaborative Research Center, IIS, University of Tokyo, Tokyo
153-8505, Japan}

\author{K.\ Sanaka}
\affiliation{Edward L. Ginzton Laboratory, Stanford University,
Stanford, California 94305-4088, USA} \affiliation{Nanoelectronics
Collaborative Research Center, IIS, University of Tokyo, Tokyo
153-8505, Japan}

\author{F.\ Yamaguchi}
\affiliation{Edward L. Ginzton Laboratory, Stanford University,
Stanford, California 94305-4088, USA}

\author{Kae Nemoto}
\affiliation{National Institute of Informatics, 2-1-2
Hitotsubashi, Chiyoda-ku, Tokyo 101-8430, Japan}

\author{W.\ J.\ Munro}
\affiliation{Hewlett-Packard Laboratories, Filton Road, Stoke
Gifford, Bristol BS34 8QZ, United Kingdom} \affiliation{National
Institute of Informatics, 2-1-2 Hitotsubashi, Chiyoda-ku, Tokyo
101-8430, Japan}

\author{Y.\ Yamamoto}
\affiliation{National Institute of Informatics, 2-1-2
Hitotsubashi, Chiyoda-ku, Tokyo 101-8430, Japan}
\affiliation{Edward L. Ginzton Laboratory, Stanford University,
Stanford, California 94305-4088, USA}

\begin{abstract}
We describe a quantum repeater protocol for long-distance quantum
communication. In this scheme, entanglement is created between
qubits at intermediate stations of the channel by using a weak
dispersive light-matter interaction and distributing the outgoing
bright coherent light pulses among the stations. Noisy entangled
pairs of electronic spin are then prepared with high success
probability via homodyne detection and postselection. The local
gates for entanglement purification and swapping are deterministic
and measurement-free, based upon the same coherent-light resources
and weak interactions as for the initial entanglement
distribution. Finally, the entanglement is stored in a
nuclear-spin-based quantum memory. With our system,
qubit-communication rates approaching 100~Hz over 1280~km with
fidelities near 99\% are possible for reasonable local gate
errors.
\end{abstract}


\maketitle


In a quantum repeater, long-distance entanglement is created by
distributing entangled states over sufficiently short segments of
a channel such that the noisy entangled states in each segment can
be purified and connected via entanglement swapping
\cite{Briegel98,Duer99}. The resulting entanglement between the
qubits at distant stations can then be used, for example, to
teleport quantum information \cite{Bennett93} or transmit secret
classical information \cite{Ekert91}. Existing approaches to
quantum repeaters generate entanglement using postselection with
single-photon detection \cite{Duan01,Childress1,Childress2}. In
these schemes, high-fidelity entanglement is created and the
subsequent entanglement purification is needed primarily to
compensate the degrading effect of connecting the imperfect
entangled pairs via swapping. However, due to their rather low
success probabilities in the initial entanglement distribution,
these protocols feature very low communication rates.

More efficient schemes, compatible with existing classical optical
communication networks, would involve bright multi-photon signals.
In this Letter, we propose such a scheme that operates in a regime
of modest initial fidelities, but creates entangled states at high
speed. The high rate in the generation of entangled pairs is
mainly due to the near-unit efficiencies for the homodyne
detection of bright signals, as opposed to the low efficiencies of
single-photon detectors. In our scheme, the resulting entangled
pairs will be discrete atomic qubit states, but the probe system
we use is a bright light pulse described and measured via a
continuous phase observable; hence, our quantum repeater is
``hybrid" not only because it employs matter signals and light
probes (as in other schemes), but more distinctly, by utilizing
both discrete and continuous quantum variables.


In general, in order to realize universal quantum computation or,
more relevant to us here, long-distance quantum communication, a
nonlinear element is needed for the implementation. Optically,
this nonlinear element may be introduced in at least two possible
ways. The first method uses only linear transformations, but a
measurement-induced nonlinearity \cite{KLM01}. In the second
approach, linear gates are supplemented by a weak nonlinear gate
where the nonlinearity is effectively enhanced through a
sufficiently strong probe beam \cite{Nemoto04}. Here we will apply
this concept to quantum communication by considering a hybrid
system based on optical carrier waves, electron-spin signals, and
nuclear-spin memories. In our proposal, a bright coherent ``probe"
pulse sequentially interacts with two electronic spins placed in
cavities at neighboring repeater stations. Entangled qubit pairs
will then be postselected conditioned upon the results of probe
homodyne measurements. Despite this postselection, high success
probabilities can still be achieved, thus keeping the main
advantage of our proposal over the single-photon-detection based
protocols \cite{Childress1,Childress2}. We will also avoid the
complication of purifying an atomic ensemble \cite{Duan01} and
directly distill the entanglement from several copies of noisy
entangled electronic-spin pairs.

The electronic and nuclear spin systems may be achieved, for
example, by single electrons trapped in quantum dots \cite{A} or
by neutral donor impurities in semiconductors \cite{B}. For a
sufficient interaction between the electron and the light, the
system should be placed in a cavity
resonant with the light; for the cavity, weak coupling is
sufficient, but a high value of $Q/V$ is required, where $Q$ is
the quality and $V$ is the mode-volume of the cavity
\cite{longerPRA}. The entire quantum repeater scheme proposed
here, including entanglement distribution, purification, and
swapping, will be based on the same bright coherent-light
resources and weak interactions.

\begin{figure}[tb]
\includegraphics[width=\columnwidth]{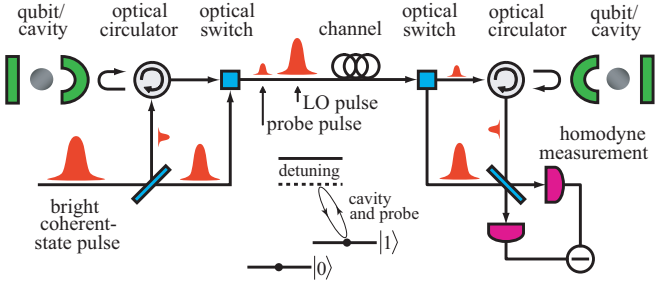}
\caption{\label{fig1} Schematic for the generation of
spin-entanglement between two qubits at neighboring stations via
homodyne detection discriminating between conditionally
phase-rotated coherent probe beams; the LO pulse is a sufficiently
strong local oscillator used for the homodyne detection.}
\end{figure}

The mechanism for the entanglement distribution among the nearest
stations of the channel is illustrated in Fig.~\ref{fig1}. The
electron spin system in the cavity is treated as a
$\Lambda$-system, with two stable or metastable ground states
$|0\rangle$ and $|1\rangle$, only one of which ($|1\rangle$)
participates in the interaction with the cavity mode. Local
rotations between states $|0\rangle$ and $|1\rangle$ may be
achieved via stimulated Raman transitions; in particular, we
suppose the state is initially prepared in the state
$(|0\rangle+|1\rangle)/\sqrt{2}$. The probe light is sufficiently
detuned from the transition between $|1\rangle$ and the excited
state to allow for a strictly dispersive light-matter interaction.
The finite probability for spontaneous emission of the qubit and
for light to leak from the cavity add a small correction to
channel losses, which we consider shortly. For clarity, let us
first discuss entanglement distribution in the absence of loss.
When the probe beam in coherent state $|\alpha\rangle$ reflects
from the cavity, the total output state may be described by
\begin{eqnarray}\label{interaction}
\hat U_{\rm int} \left[\left( |0\rangle+|1\rangle \right)
|\alpha\rangle\right]/\sqrt{2} = \left(|0\rangle
|\alpha\rangle+|1\rangle |\alpha e^{-i
\theta}\rangle\right)/\sqrt{2}\,.
\end{eqnarray}
For semiconductor impurities and realistic cavity parameters,
phase shifts of $\theta\sim 0.01$ are achievable \cite{longerPRA}.
After acquiring such a conditional phase shift at one station, the
probe beam is sent to a neighboring station and interacts with a
second spin in a similar way. Applying a further linear phase
shift of $\theta$ to the probe will yield the total state
$(\sqrt{2}|\Psi^+\rangle |\alpha\rangle + |00\rangle |\alpha
e^{i\theta}\rangle + |11\rangle |\alpha e^{-i\theta}\rangle)/2$,
where $|\Psi^+\rangle = (|01\rangle + |10\rangle)/\sqrt{2}$. Thus,
by discriminating a zero-phase shift from a $\pm\theta$ phase
shift for the probe, one can project the two spins onto a
maximally entangled state \cite{Nemoto04,Nemoto05NJoP}. Assuming
$\alpha$ real, such a projection can be approached via a $p$
quadrature measurement (i.e., along the imaginary axis in phase
space), postselecting the desired $|\Psi^+\rangle$ state.

With the ``momentum" quadrature operator defined as $\hat p =
(\hat a - \hat a^{\dagger})/2i$, the conditional state of the spin
system for a measured $p$ value of the probe beam may now be
written as
\begin{equation}
|\psi^{\text{\textsc{c}}}(p)\rangle =
\frac{C_0(p)}{\sqrt{2}}\,|\Psi^+\rangle +
\frac{C_{1}(p)}{2}\,|00\rangle + \frac{C_{-1}(p)}{2}\,|11\rangle,
\end{equation}
where $C_s(p)\equiv G_s(p)K_s(p)$, $s=0,\pm 1$, is a Gaussian
amplitude function with $G_s(p)=(2/\pi)^{1/4}\exp[-(p - s d)^2]$
and the phase factors $K_0(p)=\exp(-2 i\alpha p)$,
$K_r(p)=\exp[-i\alpha\cos\theta (2 p - r d)]$, $r=\pm 1$. In order
to assess our ability to distinguish the desired $|\Psi^+\rangle$
state (around zero phase shift of the probe) and the two unwanted
terms corresponding to the two phase-rotated probe beams, we
consider the distance of the corresponding Gaussian peaks along
the $p$ axis, $d \equiv \alpha\sin\theta$. In the following, this
parameter $d$ is referred to as the {\it distinguishability}. The
maximally entangled state is postselected by keeping the state
only when the measured result $p$ is within some finite
measurement window, $|p|<p_c$. Were it not for optical losses, a
very large window could be chosen, because by increasing $\alpha$,
the distinguishability could be made even larger, resulting in
nearly perfect postselection with probability of success $1/2$.
However, in the presence of loss, there will be a trade-off
between distinguishability and decoherence, which we now discuss.

In the presence of channel loss (and a small contribution from
cavity losses and spontaneous emission), the distinguishability
cannot be made arbitrarily large without suffering from
decoherence. We may model the photon loss by considering a beam
splitter in the channel that transmits only a part of the probe
beam with transmission $\eta^2$. The lost photons provide
``which-path" information, and tracing over them introduces the
decoherence. After the homodyne detection of the probe, the spins
are described by an unnormalized conditional density matrix
$\hat\rho^{\text{\textsc{c}}}(p)$ which depends on the measurement
result $p$ and has the following diagonal elements:
\begin{eqnarray}\label{conditionaldensitymatrix1}
\langle \Psi^\pm | \hat\rho^{\text{\textsc{c}}}(p) | \Psi^\pm
\rangle&=& |C_0(p)|^2 \,{\rm Re}(1\pm
e^{-\gamma+i\xi})/4,\\
\langle \Phi^\pm | \hat\rho^{\text{\textsc{c}}}(p) | \Phi^\pm
\rangle&=& \left(|C_{1}(p)|^2 +
|C_{-1}(p)|^2\right)/8 \nonumber\\
&&\pm e^{-\gamma}\,{\rm
Re}\left[e^{i\xi}C_1(p)\,C_{-1}^*(p)\right]/4,\nonumber
\end{eqnarray}
for the Bell states $|\Psi^\pm\rangle = (|01\rangle \pm
|10\rangle)/\sqrt{2}$ and $|\Phi^\pm\rangle = (|00\rangle
\pm|11\rangle)/\sqrt{2}$.
In the functions $C_s(p)$, $\alpha$ should now be replaced by
$\eta\alpha$ and $d$ should become $\eta d$. The decoherence in
the channel leads to a damping factor determined by
$\gamma = \alpha^2(1-\eta^2)(1-\cos\theta)\approx
\frac{1}{2}(1-\eta^2)d^2$
and an extra phase $\xi\equiv \alpha^2(1-\eta^2)\sin\theta$.

In order to maximize the distinguishability of the probe states,
we cannot simply make $d$ arbitrarily large. A correspondingly
large $d$ value would be accompanied by an increase of the
decoherence effect. This is reflected by the $d$-dependence of the
loss parameter $\gamma$. The parameter $\xi$ determines a phase
rotation, independent of the measurement result, which can be
locally removed via static phase shifters. Thus, we set $\xi\equiv
0$.

Since we cannot make $d$ arbitrarily large, we are forced to
choose a sufficiently small window for the postselection, thus
making $p_c$ sufficiently small. This will lead to a decreasing
success probability.
The success probability can
be calculated as
\begin{eqnarray}\label{probsuccess1}
P_{\rm s} = {\rm Tr}
\int_{-p_c}^{+p_c}\!dp\,\hat\rho^{\text{\textsc{c}}}(p)=
\frac{{\rm erf}(b_0)}{2} + \frac{{\rm erf}(b_1)}{4} + \frac{{\rm
erf}(b_{-1})}{4},\nonumber
\end{eqnarray}
using the diagonal elements of $\hat\rho^{\text{\textsc{c}}}(p)$
from Eq.~(\ref{conditionaldensitymatrix1}) and $b_s \equiv
\sqrt{2}(p_c + s\, \eta\, d)$, $s=0,\pm 1$. The desired entangled
output state is $|\Psi^+\rangle$, so the average fidelity after
postselection becomes
\begin{eqnarray}\label{fidelity1}
F &=& \frac{1}{P_{\rm s}}\left[\int_{-p_c}^{+p_c}\!dp\,\, \langle
\Psi^+|\hat\rho^{\text{\textsc{c}}}(p)|\Psi^+\rangle\right] \nonumber\\
&=& \frac{{\rm erf}(b_0)(1+e^{-\gamma})}{2\,{\rm erf}(b_0) + {\rm
erf}(b_1) + {\rm erf}(b_{-1})}\,.
\end{eqnarray}

\begin{figure}[t]
\includegraphics[width=8cm]{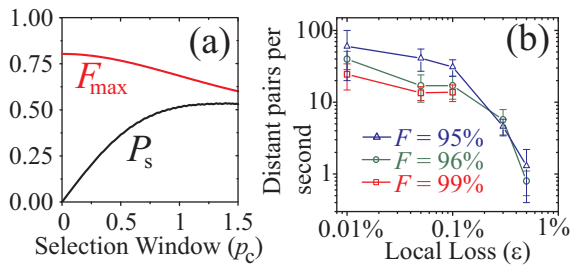}
\caption{\label{fig2} (a) the maximum fidelity $F_{\text{max}}$
and success probability $P_{\rm s}$ as a function of the
postselection window $p_c$ for
$\eta^2 = 2/3$.
(b) achievable qubit rates for different target fidelities vs.
local optical losses $\epsilon$. Each point corresponds to a
single Monte-Carlo simulation of the nested purification protocol
over 9 complete qubit teleportations; each point is the average
difference in time between teleported qubit arrival times, and the
error bar is the standard deviation.}
\end{figure}

Channel loss may come from a variety of sources, including finite
mode-coupling efficiency, but is likely to be dominated by fiber
loss. Here, we will consider only the fiber loss \cite{footnote5}.
A reasonable length for the individual segments of the quantum
repeater would be $\ell=10$~km. Assuming telecom fiber and
wavelength, where losses are about 0.17 ${\rm dB}/{\rm km}$, the
transmission parameter for 10~km is $\eta^2=2/3$. In
Fig.~\ref{fig2}(a),
the maximum fidelity
is shown as a function of $p_c$ for a transmission of $\eta^2 =
2/3$. Due to the trade-off between distinguishability and
decoherence, there is an optimal $d$-value yielding this maximum
fidelity for each $p_c$. The overall maximum fidelity of $F\approx
0.8$ for $p_c\to 0$ can be achieved only at the expense of a
vanishing success probability. However, by choosing a
postselection window $p_c\approx 0.5$ and sacrificing some
fidelity, $F\approx 0.77$, we can attain a reasonable success
probability of $P_{\rm s}\approx 36\%$. This high rate of
successful entanglement generation in our scheme is in sharp
contrast to the low efficiencies of single-photon-based approaches
\cite{Childress1,Childress2}. The above values for fidelity and
success probability correspond to distinguishabilities of $d\sim
1$, which for phase shifts of $\theta\sim 10^{-2}$ are achievable
via reasonable probe photon numbers of about $10^4$.


Our initial fidelities, $F_{\rm initial}\approx 0.77$, will be
insufficient for entanglement swapping; some entanglement
purification must first occur. For both purification and swapping,
local two-qubit gates are needed. For this purpose, we propose to
use a measurement-free deterministic controlled-phase gate based
upon a sequence of conditional rotations and unconditional
displacements of a coherent-state probe interacting with the two
spins \cite{Spiller}. The total unitary operator to achieve this
gate can be described by
\begin{eqnarray}\label{cphasegate}
\hat U_2(\theta)\hat D(\beta)\hat U_1(\theta) \hat D(-\beta^*)
\hat U_2(\theta)\hat D(-\beta)\hat U_1(\theta)\,.
\end{eqnarray}
Here, $\hat U_k(\theta)$ corresponds to the interaction in
Eq.~(\ref{interaction}), leading to a controlled phase shift of
the probe conditioned upon the state of the $k$th qubit. The
operator $\hat D(\beta)=\exp(\beta\hat a^\dagger-\beta^*\hat a)$
describes a phase-space displacement of the probe by
$\beta\equiv\alpha(1-i)$. These gate operations can be implemented
using the same bright coherent-light resources and weak
interactions as employed in the above entanglement distribution
protocol. After the entire sequence in Eq.~(\ref{cphasegate}), the
probe will be nearly disentangled from the spins. After tracing
over the probe and removing single-qubit $Z$-rotations, the qubits
have undergone a controlled phase shift of $\phi\approx
T(1+T)\alpha^2\theta^2$, where $T$ is the transmission for the
cavity-probe interaction. For a desired phase shift of the order
of $\pi$, we must satisfy $\alpha^2\theta^2 \approx 1$, which is
exactly the regime we have been using for the entanglement
distribution. For any finite optical loss, some decoherence will
occur at order $\epsilon=1-T$. A small amount of decoherence is
also introduced due to the finite probe-qubit entanglement,
scaling as $\theta^2$ if loss is neglected. The details of these
decoherence mechanisms will be discussed elsewhere
\cite{longerPRA}.

This controlled-sign gate, in addition to single-qubit rotations
and measurements (which may also be done by homodyne detection of
a bright optical probe), are sufficient resources for the standard
purification protocol introduced in Ref.~\onlinecite{Deutsch}.
This protocol was analyzed in terms of density matrices $\hat\rho$
that are exactly diagonal in the Bell-state basis. The
$\hat\rho^{\text{\textsc{c}}}$ described by
Eq.~(\ref{conditionaldensitymatrix1}) is very nearly so, and the
small off-diagonal elements quickly vanish after a few
purification steps. It was previously noted \cite{Duer99} that
this protocol converges faster than protocols based on Werner
states~\cite{Bennett96}.

Several protocols for combining entanglement purification and
swapping to connect large distances have been previously
considered. At one extreme in the number of qubit resources is
``scheme B" of D\"ur et al.~\cite{Duer99}. This scheme uses as
many qubits as are needed to allow rapid parallel purification;
for example, for communication over 1000~km, hundreds of qubits
are needed in each repeater station. At the other extreme is the
scheme of Childress~et al.~\cite{Childress1,Childress2} requiring
only two qubits per station. However, in this scheme, the
purification and swapping are very slow and become impossible if
the initial pair-fidelities are too low and gate errors are too
high.

We consider a protocol in between these two extremes; we find that
for a number of qubits per station which grows only
logarithmically with distance, a reasonable communication
qubit-rate may be achieved for reasonable gate errors. In this
scheme, each repeater station acts autonomously according to a
simple set of rules. Throughout the protocol, all stations
containing unentangled qubits simultaneously send pulses in order
to immediately attempt entanglement distribution at
nearest-neighbor stations. Meanwhile, all entangled qubit pairs
are purified a predetermined number of steps, and once this is
complete, entanglement swapping occurs to progressively double the
distance over which pairs are entangled. After entanglement
swapping, purification is again attempted, always simultaneous
with new entanglement creation at the free qubits. The limiting
timescale for these operations is the time for light to propagate
between stations in order to transmit both the entangling pulses
and the classical signals containing measurement results for
entanglement postselection, purification, and swapping.

This scheme is similar to ``scheme C" of Ref.~\onlinecite{Duer99},
where the maximum number of qubits needed per station was shown to
be $2 \log_2(\mathcal{L}/\ell)$. Here $\mathcal{L}$ is the total
length of the channel and $\ell$ is the distance between stations.
However, because of the added purification needed prior to any
entanglement swapping, we require at least $N=
2+2\log_2(\mathcal{L}/\ell)$ qubits per station. We also find that
the probabilistic creation of initial entangled pairs proceeds
more quickly if $N$ qubits are present at \emph{every} station.


As examples, we have run Monte Carlo simulations for communication
over 1280~km with repeater stations separated by 10~km, in which
case we assume 16 qubits per station. We try several choices for
the number of purification steps before and after each
entanglement-swapping step. If more purification steps are used,
larger fidelities are possible at slower rates, while fewer
purification steps lead to faster rates at smaller final
fidelities. Both the rates and fidelities drop due to local gate
errors. For our simulations we presume that these errors are
dominated by local optical loss. Figure~\ref{fig2}(b) shows
typical rates for different target fidelities and different
amounts of local optical loss.

Two more technical issues should be
raised. First, the time for optical information to propagate over
1280~km in optical fibers, about 6~ms, is already longer than
decoherence times observed in most solid-state electronic spin
systems; to this one must add the extra time required to await the
entanglement purification and swapping.
A feasible solution is the introduction of nuclear memory,
as decoherence times for isolated nuclei are at least many seconds
\cite{nuclearmemory}. For isolated nuclei, fast ENDOR
(electron-nuclear double resonance) pulse techniques may be
employed for rapid storage and retrieval of the electron-spin
state \cite{wrachtrup}. Nuclear ensembles in quantum dots have
also been considered \cite{taylormarcuslukin}, but in this case
the decoupling-limited memory time is likely to be shorter. The
second technical consideration is that for the loss-rates over the
long-distance communication channel, we have assumed telecom
wavelengths, while the solid-state emitter is likely to operate at
shorter wavelengths. Hence efficient phase-preserving wavelength
conversion of the strong optical probe is required
\cite{Langrock}.

In summary, we proposed a full quantum repeater system based upon
weak dispersive light-matter interactions. In our proposal, bright
coherent light pulses interact with small numbers of solid-state
electronic spin qubits in microcavities. The measured light
observable is a continuous phase as opposed to a discrete
occupation number. Thus, interferometric requirements are less
stringent than in many other proposals and good phase
stabilization is readily available from a phase-reference pulse
traveling down the same fiber. For the resulting high detection
efficiencies and modest initial fidelities, long-distance qubit
communication rates approaching 100~Hz with final fidelities of
99\% are possible.


The authors thank Lily Childress for useful discussions. This work
was supported in part by the JSPS, MIC, Asahi-Glass research
grants, the EU project QAP, JST SORST, ``IT  Program" MEXT, and
MURI grant\# ARMY, DAAD 19-03-1-0199.





\begin{thebibliography}{99}
\bibitem{footnote5} Other sources of loss such as finite
fiber-cavity coupling efficiencies could also be included into our
loss analysis, because modeling loss via a single transmission
parameter $\eta^2$ and tracing out lost photons will always yield
lower bounds on the achievable fidelities.
\end{thebibliography}
\end{document}